# Numerical modeling of a high power terahertz source at Shanghai [*]

DAI Jin-Hua(戴进华) [1, 2]   DENG Hai-Xiao(邓海啸) [1, 1)]   DAI Zhi-Min(戴志敏) [1]

1(Shanghai Institute of Applied Physics, the Chinese Academy of Sciences, Shanghai, 201800; China)
2 (Graduate University of the Chinese Academy of Sciences, Beijing 100049, China)

**Abstract**  On the basis of an energy-recovery linac, a terahertz (THz) source with kilowatts average power is proposed in Shanghai, which will serve as an effective tool in material and biological sciences. In this paper, the physical design of two free electron laser (FEL) oscillators, in the frequency range of 2~10THz and 0.5~2THz respectively, are presented. By using three dimensional, time-dependent numerical modeling of GENESIS in combination with paraxial optical propagation code (OPC), the THz oscillator performances, the detuning effects, and the tolerance requirements on the electron beam, the undulator filed and the cavity alignment are given.

## 1. Introduction

Terahertz (THz) wave is a frontier area for research in physics, chemistry, medicine, and biological and material sciences. THz sources of high quality have been scarce, but this gap has recently begun to be filled by a wide range of new technologies. THz wave is now available in both the continuous wave (CW) and the pulsed form, down to single-cycles or less, with the peak powers up to megawatts (MW). THz sources have led to new science in many areas, as scientists are aware of the opportunities for research progress in their fields by using THz wave. Among all THz sources, free electron laser (FEL) based on energy-recovery linac (ERL) is of great interest. The advantage of FEL is the promising prospect in high peak power, high average power and flexible wavelength tuning. At present, the Novo-FEL in BINP Russia is the most powerful THz source of the word, with a record of 0.5kW average power [1].

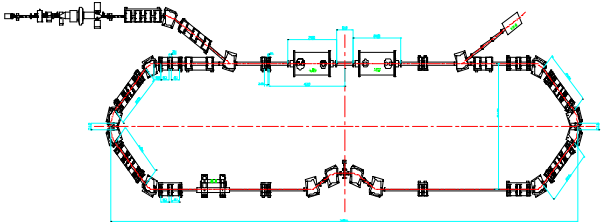

Fig. 1. Layout of Shanghai THz source based on ERL-FEL.

Recently, on the basis of ERL-based FEL oscillator, a THz source with kilowatts average power is proposed in Shanghai, China. The layout of the THz source is shown in Fig. 1. The 500keV electron bunch extracted from a 250MHz VHF gun is boosted to 2MeV at the exit of the injector. The electron beam is accelerated to 20MeV by 2 5-cell, 500MHz superconductive radiofrequency (SRF) module in the ring and transported through the THz oscillator where the kinetic energy of the electron beam is transferred to the THz radiation. The return electron bunch passes the SRF module again with a decelerating phase for energy-recovery, and then dumped. In order to achieve an average output power of 1kW, the average beam current is expected to be 20mA. The parameters of the electron beam and the details of THz oscillator are summarized in Table 1. To cover the band of 0.5~10THz, two FEL oscillators are supposed. It is 2~10THz FEL oscillator with 20MeV electron beam and 0.5~2THz with 10MeV electron beam, respectively.

Table 1. Main parameters of the Shanghai THz source

| THz frequency | 0.5~2THz | 2~10THz |
|---|---|---|
| Beam energy | 10MeV | 20MeV |
| Beam transverse emittance | 10μm-rad | 10μm-rad |
| Beam energy spread | 0.2% | 0.2% |
| Bunch charge | 240pC | 240pC |
| Bunch length FWHM | 16ps | 8ps |
| Micro pulse repetition | 83.33MHz | 83.33MHz |
| Undulator type | Helical | Planar |
| Undulator period length | 100mm | 60mm |
| Undulator periods number | 30 | 30 |
| Cavity length | 18m | 18m |
| Mirror radius of curvature | 9.2502m | 9.1160m |
| Cavity stability $g^2$ | 0.89 | 0.95 |
| Mirror radius | 120mm | 50mm |
| Out-coupling hole radius | 9.0mm | 5.5mm |
| Average output power | ~1kW | ~1kW |

This paper will present the physical optimization and numerical simulation of two FEL oscillators in the range of 2~10THz and 0.5~2THz, respectively. The numerical modeling has been carried out using three dimensional, time-dependent FEL code GENESIS [2] in combination with paraxial optical propagation code (OPC) [3]. In this paper, taking 10THz as a typical radiation wavelength, we first describe the physical design and optimization, time-dependent simulation, and tolerance performance of 2~10THz oscillator in Section 2, 3 and 4. Primary results of 0.5~2THz oscillator is illustrated in Section 5. Finally, we present our conclusions in Section 6.

## 2. THz oscillator design and optimization

The goal of a FEL oscillator design and optimization is maximum gain and optimum output power. Details of the optimization are given in this section. The cavity is of a symmetric near-concentric design with a length of 18m. The reasons of such a long cavity length choice are that, on one hand, it is helpful for expanding the radiation size of 10kW order intra-cavity power on the cavity mirror. On the other hand, a long cavity length would allow the FEL oscillator to start up when the repetition rate of the quasi-CW electron bunch decreases from 83.33MHz to 41.67, 16.67 and even 8.33MHz.

---

[*] Supported by the Chinese Academy of Science under Grant No.29Y029011
1) E-mail: denghaixiao@sinap.ac.cn

The cavity mirror radius of curvature is designed to be 9.1160m, which results in a Rayleigh length of 1.0m and a cavity stability parameter of 0.95. The 10THz optical mode radius at the undulator centre is about 3.09mm. At this location the matched electron beam radii are 0.45mm in horizontal and 0.24mm in vertical. It means that the coupling between the electron bunch and the radiation is near-optimal in a large range. Thus, the FEL oscillator performance is expected not to be degraded too much by the electron beam jitter in transverse position, which is well demonstrated in the following tolerance studies. The 10THz radiation radius on the mirror surface, assuming the fundamental transverse mode is 14.3mm compared to a mirror aperture radius about 50mm. It is large enough to ensure that the diffraction losses from the fundamental mode are minimal. All higher order modes will suffer more diffraction loss due to much wider transverse mode size, thus, limiting the mirror aperture actually is a crude method to control the transverse lasing mode.

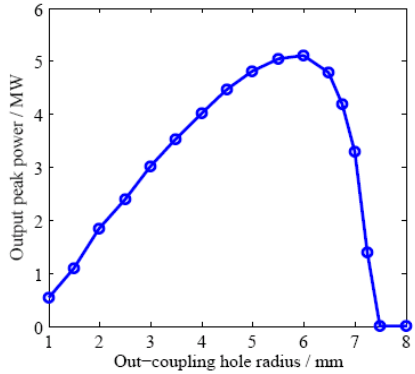

Fig 2. The output peak power as a function of the out-coupling hole radius for the 10THz oscillator.

The proposed working point of the cavity geometry shows a single-pass FEL gain larger than 50%. In order to obtain an optimum output power, we keep the mirror radius of curvature constant and change the values of the out-coupling hole radius on the upstream mirror. Fig. 2 shows the effect of varying the radius of the out-coupling hole the cavity mirrors on the output power at saturation. This work was carried out using GENESIS and OPC in steady-state mode. The results show that a coupling hole with radius of 5.5mm is near optimum in terms of the output peak power, giving an 11% out-coupling fraction. The optimal peak power is about 5MW, which agrees exactly with the theoretical power efficiency of the FEL low-gain oscillator.

## 3. Time-dependent Simulation

The time-dependent simulation allows us to model the effects of cavity length detuning, and the temporal and the spectral performances of THz oscillator in further. Using GENESIS time-dependent mode, we are however, limited to detune the cavity length only to half-integer multiples of the THz wavelength [4]. Thus, an external MATLAB script is used for coupling GENESIS and OPC, where the temporal distribution of the electron beam can be detuned by an arbitrary step with respect to the THz radiation after each FEL round trip. In this case, we are able to model the fine detuning of the cavity length.

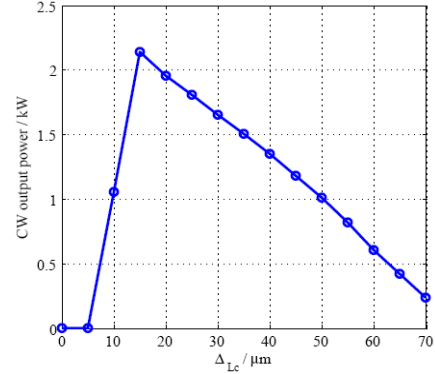

Fig 3. The detuning effect of the cavity length of 10THz FEL.

Fig. 3 plots the cavity length detuning curve of 10THz oscillator. The average output power of 10THz radiation exceeds 2kW with an optimal detuning length of 15μm. If the cavity length is effectively detuned, the evolution of the single pulse energy of the 10THz oscillator is shown in Fig. 4. It indicates 25.7μJ output pulse energy after 200 passes.

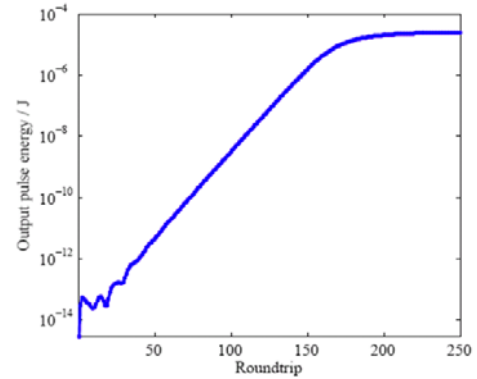

Fig 4. The pulse energy growth of the output 10THz radiation in 2~10THz oscillator.

Fig. 5 shows the temporal and spectral distribution of the 10THz output radiation at saturation with the optimal detuning length. The peak power of the 10THz radiation pulse is 5.3MW. Fig. 5 demonstrates a temporal FWHM width of 4.4ps and a spectral FWHM bandwidth of 1.5%. This corresponds to a time bandwidth product of 0.66, which is close the Fourier transform limit of 0.44 for a Gaussian pulse profile.

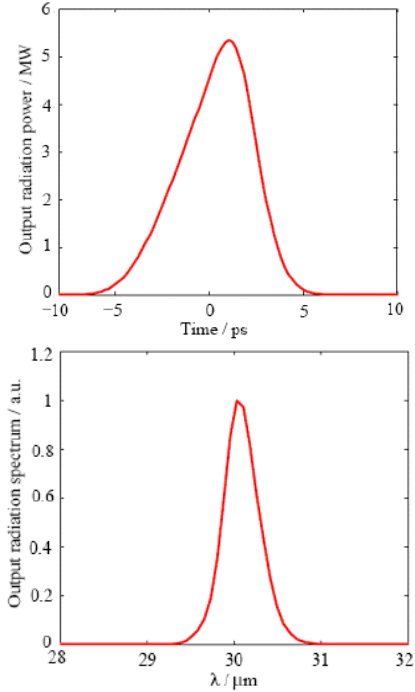

Fig. 5. The saturated output 10THz radiation pulse in time and spectral domain, where the optimal detuning is assumed.

The main effect on the electron beam due to the FEL interaction is an induced energy spread. For the 20MeV electron beam with a RMS input energy spread of 0.2%, the RMS energy spread increases to 1.3% due to the FEL interaction. Thus the exhaust energy acceptance of the beam transport return arc should be up to 8%.

## 4. Sensitivity to parameters variation

In FEL oscillator, the degradation in any parameter of the electron beam, the undulator or the THz cavity mirror could lead to a reduction of FEL performance, especially the saturated FEL output power. In this section, we will check the influence of the energy spread、the undulator field error and the cavity mirror misalignments to 10THz FEL oscillator.

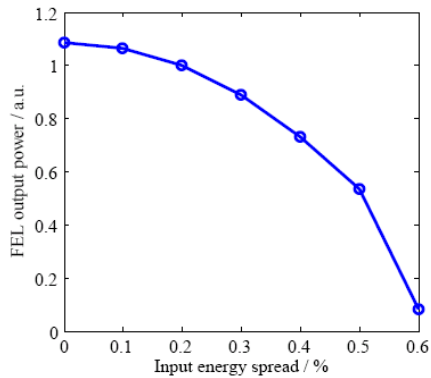

Fig 6. Normalized FEL output power .vs. initial energy spread.

For the low-gain THz oscillator, transverse emittance of the electron beam is not an issue. However, the energy spread is crucial. Fig. 6 plots the sensitivity of the FEL output power to the energy spread. If the initial energy spread increases to about 3 times of the nominal value 0.02%, the output power decreases monotonically to 8% of that produced by an electron beam with the nominal parameters. In order to assure that the THz output power could be up to 1kW, the beam energy spread should be less than 0.4%.

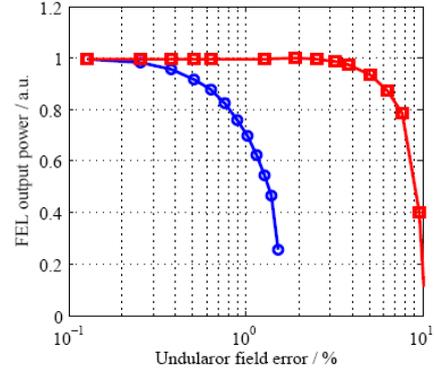

Fig 7: Normalized FEL output power .vs. undulator field errors

Undulator filed errors contribute to aberration of the beam trajectory and mismatch of the FEL resonance. Fig. 7 shows the effects of undulator errors to the FEL output power. The square line represents the case in which the beam trajectory aberration induced by the undulator field errors can be neglected, e. g. a 10% undulator field error just induce a maximum aberration of 180μm. In this case, the tolerance of undulator error is prettily relaxed. The circular line means the case in which the beam trajectory aberration induced by the undulator field errors is very serious, e. g. a 1% RMS undulator field error induce a maximum aberration of 2mm. In this case, the tolerance of RMS undulator field error should be less than 1%. Moreover, since the 10THz optical mode size is much larger than the electron beam in the undulator, the FEL performance is not sensitive to the beam centroid offset.

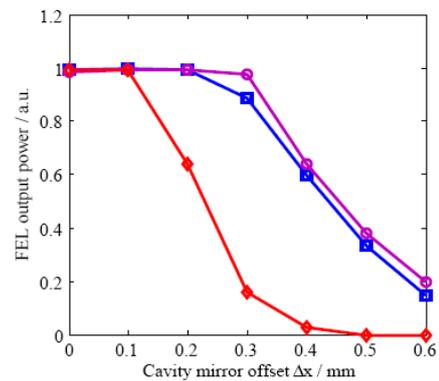

Fig 8. Normalized FEL output power .vs. cavity mirror offset.

The direct effect of the cavity misalignments is the disruption the routine propagation of the optical mode. Then it would result in the degradation of the transverse coupling between the electrons and the THz radiation, increase of the cavity loss and change of the out-coupling efficiency. Thus, the alignment of the cavity mirror is of great importance for high power THz oscillators. Fig. 8 illustrates the dependence of the FEL output power on the cavity position misalignment. The square shows the downstream mirror position offset, which indicates that the offset of the downstream mirror should be less than 0.3mm for an optimal output. The upstream mirror offset represented by the circular line shows an almost similar tendency as the downstream one. What needed to be stressed here is that, the output coupling hole is on the upstream mirror center. A small offset of the upstream mirror will induce an out-coupling position shift with respect to the optical axis of the fundamental mode. Then the disruption to the fundamental optical mode by the hole output coupling is less serious, and thus the output efficiency is enhanced when compared with a center hole coupling. This is the reason why the output power in the case with upstream mirror offset is a little bit higher than that of with downstream mirror offset. The diamond line shows the case in which the position of the two mirrors are simultaneously shifted in an opposite direction. The requirement of 0.1mm order mirror alignment can be easily accomplished by using a dedicated He-Ne laser.

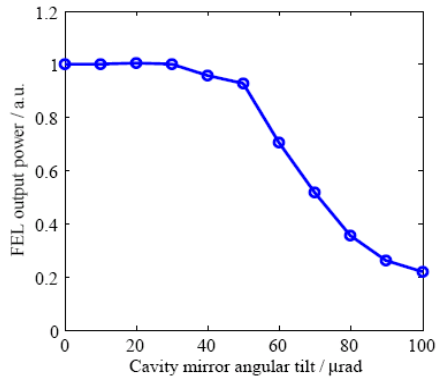

Fig 9. Normalized FEL output power .vs. cavity mirror angular misalignment.

The theory of the optical cavity predicts that the cavity mirror angular tolerance should be far less than 78μrad. Fig. 9 plots the simulated dependence of the FEL output power on the angular tilt of cavity mirror. It shows that the cavity mirror angular misalignment should be well controlled in the range of 0~50μrad, which is consistent with theoretical estimate.

## 5. Performance of 0.5~2THz oscillator

The design strategy of the 0.5~2THz mode is similar with the 2~10THz one, except that a helical undulator is employed to compensate the gain degradation caused by the strong slippage effects in long wavelength region. In the 0.5~2THz oscillator, the optimized results of 1THz radiation is given in this section.

Fig. 10 plots the cavity length detuning curve of 1THz oscillator. Within a 100μm range of the cavity detuning, the average power of the 1THz radiation is 1kW above. The average output power reaches 1.47kW with detuning length of 50μm. Fig. 11 shows the evolution of the single pulse energy of the 1THz oscillator in the case of optimal detuning. It indicates a 17.7μJ output pulse energy after 270 passes. In the operation, it should be based on the experiments to determine the specific detuning to acquire the maximum output power.

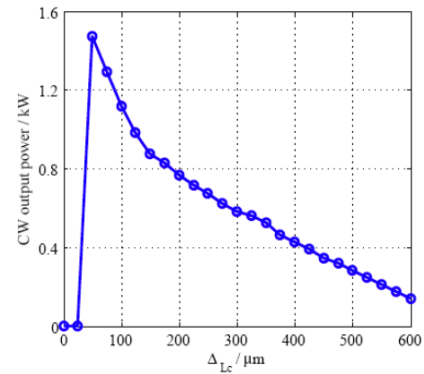

Fig 10. The detuning effect of the cavity length of 1THz FEL.

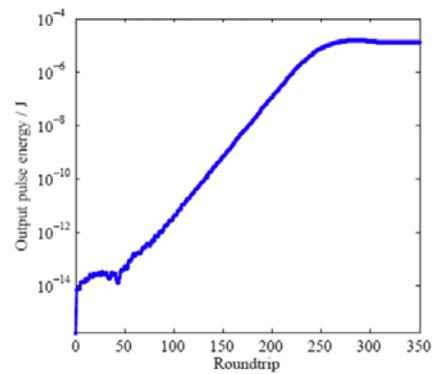

Fig 11. The pulse energy growth of the output 1THz radiation in 0.5~2THz oscillator.

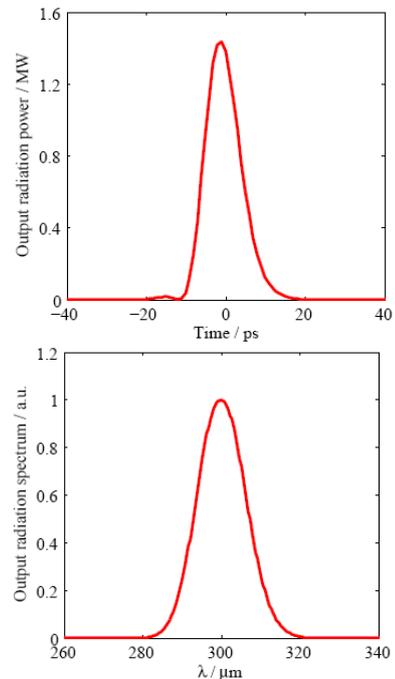

Fig. 12. The saturated output 1THz radiation pulse in time and spectral domain, where the optimal detuning is assumed.

Fig. 12 shows the temporal and spectral distribution of the 1THz output radiation at saturation with the optimal detuning length. The peak power of the 1THz radiation pulse is 1.4MW. Fig. 12 presents a temporal FWHM width of 10.2ps and a spectral FWHM bandwidth of 4.83%. This corresponds to a time bandwidth product of 0.49, which is almost the Fourier transform limit of 0.44 for a Gaussian pulse profile.

## 6. Conclusions

The ERL-based, low-gain FEL oscillator is the most attractive scheme to generate high power radiations, on the basis of which, a high power THz source with quasi CW output power about 1kW was proposed at Shanghai. The physical design and full three dimensional numerical modeling have been carried out. The result shows that, with a 20MeV, 20mA electron beam, coherent radiations with peak power of megawatts and with average power of kilowatts can be achieved in the frequency range of 0.5~10THz.

*The authors would like to thank B. Liu and P. J. M. Vander slot for help in executing OPC code.*